\documentclass[prb, twocolumn, superscriptaddress, longbibliography]{revtex4-2}

\usepackage{graphicx} 
\usepackage{enumitem}
\usepackage{float}
\usepackage{physics}
\usepackage{amsmath}
\usepackage{balance}
\usepackage{color}
\usepackage{lastpage}
\newcommand{\red}{\textcolor{black}}

\usepackage[colorlinks=True]{hyperref}  
\hypersetup{
 	colorlinks=true,  
 	linkcolor=blue,  
 	citecolor=blue, 
 	filecolor=magenta,   
 	urlcolor=blue         
 }

\usepackage{xcolor}

\makeatletter
\def\maketitle{
\@author@finish
\title@column\titleblock@produce
\suppressfloats[t]}
\makeatother

\makeatletter
\newcommand\footnoteref[1]{\protected@xdef\@thefnmark{\ref{#1}}\@footnotemark}
\makeatother

\makeatletter
    \def\balanceissued{unbalanced}
    \let\oldbibitem\bibitem
    \def\bibitem{%
        \ifnum\thepage=6%
            \expandafter\ifx\expandafter\relax\balanceissued\relax\else%
                \balance%
                \gdef\balanceissued{\relax}\fi%
            \else\fi%
        \oldbibitem}
\makeatother

\begin{document}


\title{Sharp Page transitions in generic Hamiltonian dynamics}

\author{Lauren H. Li} 
\affiliation{Department of Physics, Princeton University, Princeton, New Jersey 08540, USA}

\author{Stefan Kehrein}
\affiliation{Institute for Theoretical Physics, Georg-August-Universität Göttingen, Friedrich-Hund-Platz 1, 37077 Göttingen, Germany}

\author{Sarang Gopalakrishnan}
\affiliation{Department of Electrical and Computer Engineering, Princeton University, Princeton, New Jersey 08540, USA}

\date{July 8, 2026}

\begin{abstract}
We consider the entanglement dynamics of a subsystem initialized in a pure state at high energy density (corresponding to negative temperature) and coupled to a cold bath. The subsystem's R\'enyi entropies $S_\alpha$ first rise as the subsystem gets entangled with the bath and then fall as the subsystem cools. We find that the peak of the min-entropy, $\lim_{\alpha \to \infty} S_\alpha$, sharpens to a cusp in the thermodynamic limit at a well-defined time we call the Page time. 
We construct a hydrodynamic ansatz for the evolution of the entanglement Hamiltonian, which accounts for the sharp Page transition as well as the intricate dynamics of the entanglement spectrum before the Page time. 
%
Our results hold both when the bath has the same Hamiltonian as the system and when the bath is taken to be Markovian. Our ansatz suggests conditions under which the Page transition should remain sharp even for R\'enyi entropies of finite index $\alpha$.
%

\end{abstract}

\maketitle
\section{Introduction}

As a black hole evaporates, its entanglement structure undergoes a qualitative change. At early times, the black hole is only weakly entangled with its surroundings. Emitting Hawking radiation entangles the degrees of freedom in the black hole with those asymptotically far away. Once more than half the black hole has evaporated, its remaining degrees of freedom become maximally entangled with previously emitted (early) radiation; subsequent (late) radiation can only decrease the black hole's entanglement entropy. The time at which the black hole becomes maximally entangled with its surroundings is called the Page time~\cite{polchinski2017black}. It follows from monogamy of entanglement that radiation emitted after the Page time is unentangled with the remaining black hole. This lack of entanglement poses a conceptual paradox since, from the perspective of the effective field theory of semiclassical gravity, nothing changes qualitatively at the Page time~\cite{almheiri2013black, polchinski2017black, susskind2013black, harlow2013quantum}. 
If the effective field theory offers a consistent description of at least \emph{some} observables, the singular change in the entanglement structure of the black hole must be hidden from these observables. 
Recent proposals to reconcile semiclassical gravity and quantum mechanics have drawn on ideas from quantum information theory, such as non-isometric codes~\cite{akers2024black, kim2023complementarity, dewolfe2023non, PhysRevD.109.044005} and cryptographic censorship~\cite{engelhardt2024cryptographic}. Since these proposals are not inherently tied to gravity, it is natural to explore realizations of the Page transition and its associated paradoxes in the simpler context of non-relativistic many-body physics.

Indeed, Page's original argument does not invoke gravity or even dynamics~\cite{page2013time}. Instead, it models the black hole and its emitted radiation as collectively forming a random pure state on $N$ qubits. Each qubit is randomly assigned to subsystem $A$ (the black hole) or its complement $\overline{A}$ (the radiation). The age of the black hole is modeled by the number of qubits remaining in $A$, denoted $N_A$. A simple calculation shows that, up to subleading corrections, the entanglement entropy of the black hole follows $S = \min(N_A, N_{\overline{A}})$ in the limit $N_A, N_{\overline{A}} \to \infty$. 
If one thinks of the entanglement entropy as a high-temperature thermodynamic entropy, $S$ can be said to undergo a \emph{first-order} transition.
Page's toy model assumes that the black hole scrambles information instantaneously; whether Page-like transitions remain sharp under more realistic local quantum dynamics remains an open question (for recent discussion, see Refs.~\cite{PhysRevD.100.105010, blake2023page, PhysRevD.109.L081901, PhysRevD.102.086017, piroli2020random, PhysRevB.109.224308,  saha2024generalized, glatthard2025thermodynamics, ganguly2025quantum}).


As one of us recently showed~\cite{PhysRevB.109.224308}, this question has an affirmative answer for dynamics with a conserved charge. 
To realize a Page curve, we consider a universe initially in the vacuum state of this charge, except for a region $A$ (analogous to the black hole) with a high charge density. We take $A$ to be much smaller than its complement $\overline{A}$. Under time evolution, charge spreads from $A$ to $\overline{A}$. In the early stages of this spreading, the entanglement entropy of $A$ increases; however, at very late times, $A$ will have a low charge density above the vacuum, limiting its effective Hilbert space dimension and, hence, the amount of entanglement it can support. Consequently, the entanglement must evolve non-monotonically, reaching a maximum at some intermediate time. This non-monotonicity does not necessarily imply non-analytic behavior in the entanglement evolution. In fact, in the explicitly solvable case of free fermions~\cite{PhysRevB.109.224308}, all the R\'enyi entanglement entropies, $S_\alpha = (1-\alpha)^{-1} \mathrm{Tr}(\log \rho_A^\alpha)$, where $\rho_A$ is the reduced density matrix of region $A$, evolve analytically except in the $\alpha \to \infty$ limit. Intriguingly, this quantity, known as the min-entropy $S_\infty \equiv -\log \lambda_{\mathrm{max}}$, where $\lambda_{\mathrm{max}}$ is the largest eigenvalue of $\rho_A$, exhibits multiple cusps even before the entanglement reaches its maximum. These cusps have a simple origin: for charge-definite initial states, $\rho_A$ is block-diagonal, with each block corresponding to a definite value of the total charge in region $A$. As the typical charge in $A$ evolves, eigenvalues in distinct blocks can cross, and these crossings remain sharp even in finite systems. 
In the thermodynamic limit, the crossings merge to form a continuum, leaving only the first crossing responsible for non-analytic behavior. These findings for a non-interacting, analytically solvable model \cite{PhysRevB.109.224308} have recently been extended to interacting models with a conserved charge \cite{RJha2025}.

In this work, we explore the Page transition for chaotic quantum dynamics where energy is the only conserved quantity. The Hamiltonians we consider will be geometrically local; that is, we can write $H = \sum_x h_x$, where each term $h_x$ is supported on finitely many contiguous sites. We briefly comment at the end on the case of long-range interactions. 
We consider initial states that are in or near the ground state everywhere except in  region $A$. Typically, we will initialize region $A$ at an energy density corresponding to negative temperature and study the evolution of entanglement between $A$ and $\overline{A}$, described by the time-evolving density matrix $\rho_A(t)$. In addition to exploring the full-system pure-state evolution, we introduce a simplified model in which $\overline{A}$ is replaced by a Markovian bath that cools the system to a low temperature. This simplification allows us to treat the evolution using a Lindblad master equation, enabling access to larger system sizes. 

We characterize entanglement dynamics through the low-energy part of the entanglement spectrum. Recall that one can generally write a density matrix $\rho_A \equiv \exp(-H_A)$, where $H_A$ is called the entanglement Hamiltonian and its spectrum is called the ``entanglement spectrum.'' The low-energy part of the entanglement spectrum corresponds to the large-weight eigenvalues of $\rho_A$. 
In general, $H_A$ need not be local, but it is known to be local for gapped ground states of local Hamiltonians~\cite{PhysRevLett.101.010504}. For highly excited eigenstates of chaotic Hamiltonians, the eigenstate thermalization hypothesis implies that the subsystem density matrix satisfies $\rho_A \propto \exp(-\beta H)$, leading to $H_A = \beta H$~\cite{RevModPhys.83.863, PhysRevX.8.021026, zhu2020entanglement}. \red{We can define a partition function as $Z\equiv\Tr\exp(-\beta H)$, from which the free energy follows as $F = -\beta^{-1}\log Z$. This naturally yields an entropy $S_{\beta}=\beta F/(\beta-1)$, which is exactly the R\'enyi entropy; the R\'enyi index $\alpha$ plays the role of an inverse temperature $\beta$. Any non-analytic behavior observed in $S_{\infty}$ hence corresponds precisely to non-analyticity in the ground state of $H_A$, and sharp transitions of $S_{\infty}$ can be thought of as quantum phase transitions.}

\red{To further explore this thermodynamic interpretation,} we numerically extract $H_A(t)$ for the time-evolving state and find that it obeys a form of local thermalization: namely, $H_A(t) \approx \sum_x \beta(x,t) h_x$, where $\beta(x,t)$ is a local effective temperature that evolves according to hydrodynamics. This provides a simple interpretation of the Page transition: it corresponds to a sign change in the entanglement Hamiltonian, where the parameters $\beta(x,t)$ pass through zero. Since the ground states of $H$ and $-H$ are macroscopically distinct, the Page transition is generally a first-order transition in the entanglement spectrum. Our numerical results support this picture: the minimum gap of $H_A$ closes roughly exponentially with system size, through an isolated level crossing between macroscopically distinct states. In addition to the Page transition, we find various well-defined crossovers in the entanglement dynamics, which correspond to the development of zeros in the ``temperature profile'' $\beta(x,t)$. We argue, and have verified, that these features are not model specific: they occur for generic high-energy initial states coupled to generic cold baths. Thus, our results suggest that the evolution of system-environment entanglement in open quantum systems can undergo singular changes that remain invisible in the dynamics of local observables.

The rest of this work is structured as follows. In Sec.~\ref{model}, we introduce the Hamiltonian and Lindbladian models that we study. In Sec.~\ref{entdyn}, we explore key features in the evolution of $H_A$ and the entanglement entropy in both models; we also present evidence that the Page transition, at least in the Lindbladian model, is sharp in the thermodynamic limit. In Sec.~\ref{hydro}, we develop a hydrodynamic picture of entanglement dynamics in terms of the parameters $\beta(x,t)$ introduced above. Finally, in Sec.~\ref{disc}, we comment on the implications of our findings, their experimental relevance, and potential extensions to higher dimensions and long-range interactions.

\section{Models}\label{model}

In this section, we introduce two classes of dynamics for which a Page transition occurs. In both cases, the analog of the black hole is a spin chain of $M$ sites, evolving under the mixed-field Ising Hamiltonian,
\begin{equation}\label{mfim}
H = g\sum_{i} \sigma_i^x + h\sum_{i} \sigma_i^z + J\sum_{i}\sigma_i^z\sigma_{i+1}^z, 
\end{equation}
for $g=0.905$, $h=0.809$, and $J=1$ to ensure the system remains robustly nonintegrable~\cite{kim2013ballistic}. We have verified that our results hold for other parameter choices, which yield qualitatively similar behavior. For simplicity, we will therefore restrict our discussion to this well-characterized parameter set.

\subsection{Full-system dynamics}

The most direct way to introduce a bath is to extend the system to $M + N$ sites, all governed by the Hamiltonian~\eqref{mfim}. The leftmost $M$ sites define region $A$, while the remaining sites form its complement $\overline{A}$ (Fig.~\ref{fig:fig1}a). 
%
%
We initialize this system of $M + N$ spins in a state that is unentangled between the regions $A$ and $\overline{A}$: 
\begin{equation}
    \ket{\psi(t=0)} = \ket{\text{ceiling}}_A \otimes \ket{\text{ground}}_{\overline{A}},
\end{equation}
where ground (ceiling) denotes the lowest (highest) energy eigenstate of $H$ in the given region. 
Under unitary evolution of the entire system, the entanglement between $A$ and $\overline{A}$ initially increases before decreasing, as expected~(Fig.~\ref{fig:fig1}c). In particular, the min-entropy exhibits a sharp peak, which we refer to as the Page time. This peak appears as long as $N \gg M$; to ensure a clear separation between the peak and the late-time entanglement, we choose $N = 3 M$. In our simulations using Krylov-space time evolution, we can efficiently study full systems of size $M + N = 24$, but larger sizes become computationally challenging. 

\subsection{Lindbladian dynamics}

\begin{figure}
    \centering
    \includegraphics{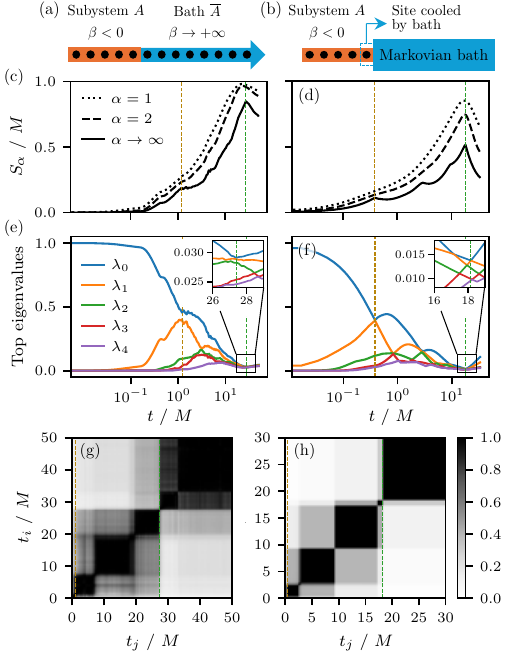}
    \vspace{-15pt}
    \caption{Existence of Page-curve in non-integrable systems. (a) The initial state consists of a subsystem of size $M$ at $\beta<0$ and bath of size $N$ at $\beta\to+\infty$; we consider the limit where $N \gg M$. (b) We also introduce a Markovian bath that cools the rightmost site. Both models exhibit entanglement entropy dynamics resembling a Page curve. (c,d)~Von Neumann entropy $S_1$, R\'enyi-2 entropy $S_2$, and min-entropy $S_{\infty}$ for full-system and Lindbladian dynamics, respectively. (e,f) The eigenvalue spectrum of both models show avoided crossings before the Page time, respectively; insets highlight the transition at the Page time. (g,h)~Overlaps between the eigenvector associated with the highest eigenvalue at $t_i$ and $t_j$ according to the Bhattacharyya distance for both models, respectively. Changes in eigenvectors resemble Landau-Zener transitions. The vertical lines mark the time of the first avoided crossing (tan) and the Page time (green). For Krylov-space time evolution, we take $M=6$, $N=18$; for Lindbladian simulations, we take $M=12$ and $\gamma=0.2$.}
    \label{fig:fig1}
\end{figure}

The system sizes reachable with exact evolution are limited by the need to track the dynamics in the bath region, $\overline{A}$. In general, this bath is non-Markovian: generic correlations in the bath are governed by energy diffusion. However, since our qualitative argument for the Page transition did not rely on the dynamics of the bath, it is natural to ask whether the same phenomenon can be observed in a model where region $A$ is coupled to a Markovian bath. We now introduce such a model and show that it exhibits entanglement dynamics similar to the full-system evolution (Fig.~\ref{fig:fig1}b). 

Specifically, we consider dynamics governed by the master equation $\partial_t \rho = \mathcal{L}(\rho)$ for the density matrix $\rho$ of subsystem $A$, where $\mathcal{L}$ is the Lindblad superoperator: 
\begin{equation}\label{lindblad0}
\mathcal{L}(\rho) = - i [H, \rho] + \gamma \sum\nolimits_k (Q_k \rho Q_k^\dagger - \frac{1}{2} \{ Q_k^\dagger Q_k, \rho \}).
\end{equation}
Here, $H$ is the mixed-field Ising Hamiltonian~[Eq.~\eqref{mfim}], the brackets $\{ \}$ denote an anticommutator, and $Q$ is a dissipative coupling acting on the two rightmost sites of region $A$. To construct $Q_k$, we first define the Hamiltonian for the two rightmost sites of $A$:
\begin{align}
    h_{M-1,M} &= g\sigma_{M-1}^x+h\sigma_{M-1}^z+g\sigma_{M}^x+h\sigma_{M}^z+\sigma_{M-1}^z\sigma_{M}^z \nonumber\\
    &=\sum_k E_k \dyad{k}{k},
\end{align}
where the second line expresses $h_{M-1, M}$ in its eigenbasis. Defining $\ket{0}$ as the ground state of the local Hamiltonian $h_{M-1, M}$, we can now write the jump operators $Q_k$ as
\begin{equation}\label{jump}
Q_k = \dyad{0}{k}.
\end{equation}
The Lindbladian dynamics is generated by the superoperator~[Eq.~\eqref{lindblad0}] with jump operators~[Eq.~\eqref{jump}]. Once again, we initialize subsystem $A$ in the ceiling state.
Just as in the non-Markovian case, we observe Page-curve-like behavior, with the entanglement entropy initially increasing then decreasing~(Fig.~\ref{fig:fig1}d). However, for the Lindbladian dynamics, the unique steady state is one in which all the jump operators are satisfied---namely, the ground state of $H$. As a result, the entanglement of $A$ decays to a much lower value than in the non-Markovian case at late times.

Some comments on this Lindbladian model are in order. First, the Lindbladian dynamics has a significant effect on the hydrodynamics of energy: instead of simply diffusing through the full system, energy now vanishes irreversibly at a sink at the right edge of subsystem $A$. Nevertheless, the dynamics in the bulk of this region is still diffusive, so it is unsurprising that $\rho$ still undergoes a Page transition. Second, because the Lindbladian~[Eq.~\eqref{lindblad0}] has only four jump operators, independent of $M$, its dynamics can be Trotterized (with a small time step, for which we set to $\delta t=0.2$) and simulated in the Kraus representation. The dominant computational cost comes from the matrix multiplication $e^{-i H t} \rho e^{i H t}$. This approach allows us to simulate systems with $M = 12$ to very late times---an infeasible task with full-system dynamics. \red{Third, because Eq.~\eqref{lindblad0} involves only a few jump operators, it can be efficiently implemented on a quantum device with $M + 2$ qubits, where the last two qubits serve as ancillas, which we assume can be reset, as in recent experiments~\cite{mi2024stable}. }


\section{Entanglement dynamics}\label{entdyn}

We begin with a qualitative discussion of the entanglement dynamics before addressing the quantitative question of whether the Page transition is sharp. The qualitative discussion applies to both the full-system and Lindbladian dynamics. However, for full-system dynamics, we are unable to study a sufficiently wide range of system sizes to address quantitative size dependence; therefore, our quantitative analysis will focus on the Lindbladian model, though we expect similar conclusions to hold for the full-system dynamics.

\subsection{Evolution of the entanglement spectrum}

We first examine how the R\'enyi entanglement entropies $S_\alpha$ of subsystem $A$ evolve. We explicitly show the cases $\alpha = 1$ (Von Neumann entropy), $\alpha = 2$ (R\'enyi-2 entropy), and $\alpha \to \infty$ (min-entropy) (Figs.~\ref{fig:fig1}c,d). All three R\'enyi entropies exhibit non-monotonic behavior, each reaching a well-defined maximum at the Page time (which coincides for all $\alpha$ to within our resolution). However, the growth of $S_\infty$ displays additional features at times preceding the Page time. Moreover, $S_\infty$ appears to form a cusp at the Page time rather than a smooth maximum. Note that in the Lindbladian dynamics, the maximum entanglement entropy occurs at lower entanglement density than in the full-system dynamics. This difference arises because, in the full-system dynamics, subsystem $A$ effectively reaches infinite temperature at the Page time. In contrast, the Lindbladian bath continuously extracts energy from the system, keeping the edge closest to the bath cold and limiting the maximum achievable entropy. \red{At late times, the state evolving under Lindbladian dynamics settles into a non-equilibrium steady state, consistent with the general results of boundary-driven open quantum systems~\cite{carlen2025stationary}. The entanglement entropies then saturate at a value that scales linearly with $M$~\cite{suppmat}.} 

To better understand the dynamics of $S_\infty$, we examine how the largest few eigenvalues of the reduced density matrix $\rho_A$ evolve. Equivalently, if one thinks of $\rho_A = \exp(-H_A)$, where $H_A$ is the entanglement Hamiltonian, we are considering the low-energy part of the entanglement spectrum. The bumps in $S_\infty$ correspond to avoided level crossings in the spectrum of $\rho_A$~(Figs.~\ref{fig:fig1}e,f). A series of such level crossings occurs as time passes; empirically, the Page transition coincides with the second-to-last and last of such level crossings involving the top eigenvalue of $\rho_A$ in the full-system and Lindbladian dynamics, respectively. These avoided level crossings correspond to discrete jumps in the nature of the top \emph{eigenvector} of $\rho_A$~(Figs.~\ref{fig:fig1}g,h) and are qualitatively similar to Landau-Zener transitions. To show that the crossing eigenvectors are supported on distinct configurations in the computational basis, we plot each eigenvector as a probability distribution over all possible states in the computational basis and find the overlaps of these distributions according to the Bhattacharyya distance~\cite{fukunaga2013introduction}:
\begin{equation}
    \mathfrak{B} = \sum_{x\in\chi}\sqrt{P(x)Q(x)},
\end{equation}
where $P(x)$ and $Q(x)$ are two probability distributions over a discrete domain $\chi$ of computational basis states. 

\subsection{Sharpness of crossings}

Given the similarity between the anticrossings in the entanglement spectrum and Landau-Zener transitions, it is natural to ask how the matrix elements for these putative transitions---specifically, the minimum gap at the anticrossing---depend on the size of $M$. The approximate crossings in Figs.~\ref{fig:fig1}e,f are pronounced enough that, particularly in the Lindbladian model, it is difficult to visually determine whether they are truly avoided. To address this question, we analyzed the dynamics of these crossings in the Lindbladian model as a function of $M$, finely sweeping parameters across each crossing to locate the minimum gap. Our results are shown in Figs.~\ref{fig:fig2}a,b for the earliest avoided crossing and the crossing that occurs at the Page time, respectively. The system-size dependence differs markedly between these two cases: for the earliest crossing, the minimum gap stabilizes to a small but clearly nonzero value; at the Page time, however, the minimum gap appears to decrease approximately exponentially with the size of $M$. This data supports the thesis that the Page transition is sharp \red{in the thermodynamic limit}, while the earlier entanglement features are crossovers. \red{In this sense, the transition resembles a quantum phase transition, characterized by a gap that only closes in the thermodynamic limit---in contrast to examples where exact crossings occur at finite sizes due to symmetries or other constraints.}

\subsection{Metastability across the Page transition}

\begin{figure}[!t]
    \centering
    \includegraphics{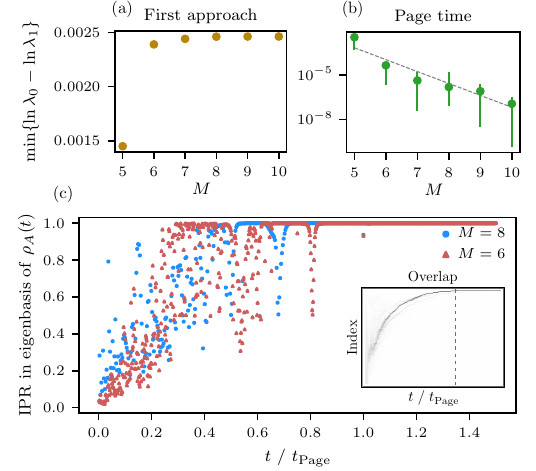}
    \vspace{-15pt}
    \caption{Minimum difference between the log of the top two eigenvalues at the (a) first approach and (b) Page time, with respect to system size. This difference converges to a finite value at the first avoided crossing, while we observe an exponential decay at the Page time. In the thermodynamic limit, we expect this difference to decrease to zero at the Page time, indicating a sharp transition. Note that a power law decay could also be consistent with the data due to the large error bars. We estimate the errors by considering the resolution of the sampling. (c) Inverse participation ratio (IPR) of the ground state of $H_A$, $\ket{\psi_0}$ (which is the dominant eigenstate of $\rho_A$ at late times) in the instantaneous eigenbasis of $\rho_A(t)$, which consists of eigenvectors $\ket{\psi_n(t)}$ with \emph{index} $n$. A large value means that $\ket{\psi_0}$ is close to being an instantaneous eigenstate of $\rho_A(t)$. Inset: Two-dimensional density plot of $|\langle \psi_0 | \psi_n(t)\rangle|$, as a function of index $n$. After the Page time, $\ket{\psi_0}$ is the ground state of the entanglement Hamiltonian; before the Page time, it is close to being a well-defined excited state of the entanglement Hamiltonian.}
    \label{fig:fig2}
\end{figure}

First-order transitions generally feature metastability, where the ground state on one side of the transition persists a long-lived resonance (or false vacuum) on the other side. Across the Page transition, metastability would imply that the ground state of the entanglement Hamiltonian at late times (which is the ground state of $H$) remains a sharp resonance in the entanglement spectrum \emph{before} the Page time (Fig.~\ref{fig:fig2}c). This is a nontrivial consequence of the Page transition being sharp. Since the state of the system prior to the Page time is far from equilibrium, there is no \emph{a priori} reason for the ground state of $H$ to be a well-defined feature in the entanglement spectrum. Nevertheless, we find that the ground state remains sharply localized in the instantaneous eigenbasis of $\rho_A$ for a significant temporal range before the Page time. These findings further support the thesis that the Page time is a first-order entanglement transition.



\section{Hydrodynamic interpretation}\label{hydro}

In this section, we provide an interpretation of these numerical findings. To motivate the discussion, we first consider a toy version of the model in which the coupling between $A$ and $\overline{A}$ is taken to be much weaker than the thermalization rate of $A$. The subsystem $A$ is initially in a thermal state at low negative temperature and eventually ends up in a thermal state at low positive temperature. In this toy model, $A$ remains in a thermal state throughout the process, so $\rho_A(t) \propto \exp(-\beta(t) H)$. At the Page time, $\beta(t)$ passes (presumably smoothly) through zero. This toy model already exhibits a phase transition in $S_\infty$, in general. For times before (after) the Page time, the top eigenvector of the density matrix is the ground state of $-H\, (+H)$. These two ground states are macroscopically distinct, so in the toy model, the Page time marks a first-order transition in $S_\infty$. In a finite system of $M$ sites, this transition would manifest itself as an avoided crossing with a minimum gap of order $O(\exp(-M))$, consistent with the numerical results of the previous section. 

Although this toy model captures some features of our numerics, it is unrealistic in the standard thermodynamic limit: the rate at which the system couples to the bath is $O(1)$, while the rate at which energy diffuses across the system is $O(1/M^2)$. Until the Page time, this disparity creates energy gradients across subsystem $A$. Consequently, the top eigenvector of $\rho_A(t)$---which is also the best rank-1 approximation to the density matrix---is spatially nonuniform through the Page time. 
The top eigenvector is clearly not the ground state of either $H$ or $-H$.

\begin{figure*}
    \centering
    \includegraphics[width=480pt]{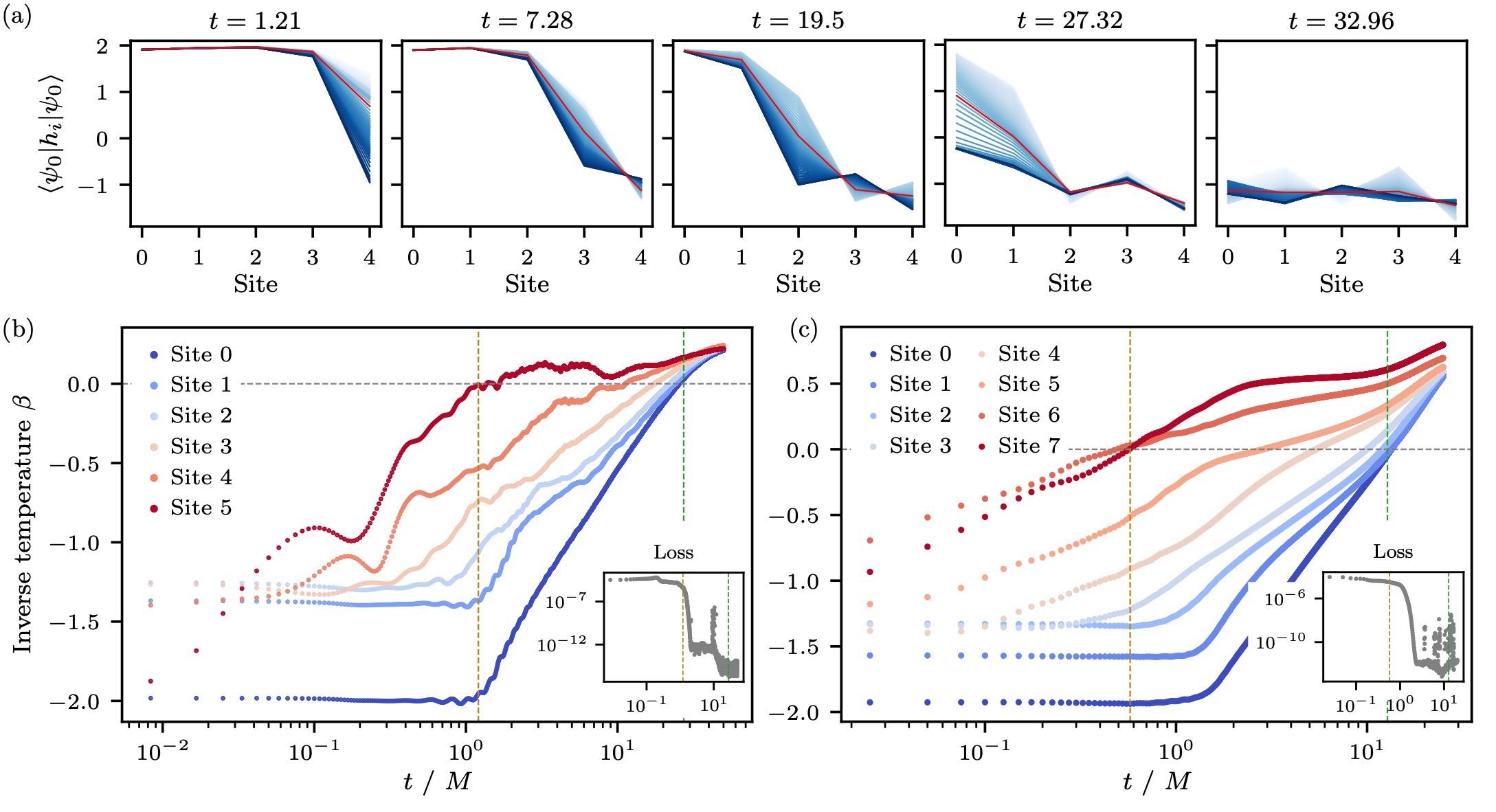}
    \vspace{-5pt}
    \caption{Hydrodynamic interpretation. (a) Energy profile of the top eigenvector of $\rho_A$ as a function of time. The red line indicates the energy profile at the given time, while the light (dark) blue lines correspond to a sequence of earlier (later) times. (b,c) Fits of the local effective temperatures [see Eq.~\eqref{rhoansatz}] for full-system simulations ($M=6$) and Lindbladian simulations ($M=8$), respectively. The vertical lines indicate the first avoided crossing and the Page transition. Note that the Page transition occurs when the effective inverse temperatures at the site farthest from the bath (site 0) changes sign. The inset shows the loss function, calculated as $\sum_i(\Tr{h_i\rho(t)}-\Tr{h_i\rho_{A}})^2$, where $\rho(t)$ is the time-evolved local density matrix and $\rho_A$ is the $\beta$-fitted ansatz.}
    \label{fig:fig3}
\end{figure*}

To move beyond the toy model described above, we assume that subsystem $A$ rapidly approaches a \emph{locally thermal} state with a spatially varying temperature profile. This assumption is standard in hydrodynamics~\cite{bulchandani2021superdiffusion} and is natural when the underlying dynamics is strongly chaotic. We express  the Hamiltonian Eq.~\eqref{mfim} as $H = \sum_i h_i$, with local energy density
\begin{equation}
    h_i = g\sigma_i^x + h\sigma_i^z + J\sigma_i^z\sigma_{i+1}^z.
\end{equation}
A natural ansatz for a locally thermal density matrix is 
\begin{equation}\label{rhoansatz}
    \rho_A(t) \propto \exp{-\int\beta(x,t)h(x) dx},
\end{equation}
where, in the large-$M$ limit, the local energy density evolves according to a diffusion equation, and the evolution of $\beta(x,t)$ is governed by energy diffusion. In practice, since we are working with small systems, we determine the profile of $\beta(x,t)$ by fitting the exact time-evolved state to the form of Eq.~\eqref{rhoansatz}. Additionally, we determine $\beta(x,t)$ separately for sites $\beta_i(t)$ and for bonds $\beta_{i,i+1}(t)$. This local thermalization ansatz ignores energy currents, which are significant at early times. To refine the ansatz, we incorporate a nonequilibrium current correction $J_i^{ZX} \sigma^z_i \sigma^x_{i+1}$. These changes lead to the replacement:
\begin{eqnarray}
&& \int dx \beta(x,t) h(x) \\
&& \quad \rightarrow \sum\nolimits_i  \beta_i(t)(g\sigma_i^x + h\sigma_i^z) + \beta_{i,i+1}(t)J\sigma_i^z\sigma_{i+1}^z \nonumber\\
&& \quad\quad\quad\quad\quad\quad + J^{ZX}_i(t)\sigma_i^z\sigma_{i+1}^x \nonumber.
\end{eqnarray}
We show in Fig.~\ref{fig:fig3}a how the local energies evolve according to the top eigenvector $\ket{\psi_0}$. The site-specific and bond-specific energies exhibit similar behavior (see Ref.~\cite{suppmat}). 

Two features stand out: the first avoided crossing occurs when the temperature at the site nearest to the bath changes sign; the last avoided crossing occurs at the Page time, when the temperature at the sites farthest from the bath changes sign (Figs.~\ref{fig:fig3}b,c). This hydrodynamic perspective naturally explains why the early-time crossings are avoided while the late-time crossings are not. The earliest crossing occurs when the preferred state of a single spin and its adjacent bonds changes. As a result, the ground state changes \emph{locally} rather than macroscopically, maintaining a finite gap between the top two eigenvalues. As time progresses, a series of such crossings occurs as spins farther from the bath cool to positive temperature. However, due to energy diffusion, temperature gradients diminish over time. Hence, later crossings involve increasingly collective processes, where many spins flip simultaneously. In this scenario, the Page time corresponds to the time required for energy to diffuse across region $A$. At the Page time, energy gradients vanish and the size of the region undergoing spin flips diverges. While this scenario applies to an infinite system, similar behavior is observed even in relatively small systems. Specifically, the zero-temperature crossings of spins far from the bath become increasingly close together and accumulate near the Page time. 

\section{Discussion}\label{disc}

In this work, we explored the Page transition under generic chaotic Hamiltonian dynamics, with energy as the only conserved quantity. Our numerical evidence suggests that the Page transition remains a sharp transition in the entanglement spectrum---specifically, a cusp in the value of the largest Schmidt coefficient---in the limit of a large subsystem $A$. We also identified a series of avoided level crossings in the entanglement spectrum prior to the Page transition and found that these crossings remain open in the thermodynamic limit. We explained these phenomena in terms of local thermalization to a state with a spatially varying effective temperature. As the effective temperature evolves, the coupling constants in the Hamiltonian change sign, leading to avoided level crossings. The Page transition corresponds to the point at which the effective temperature of the bulk of subsystem $A$ changes sign. As a by-product of this work, we found a compact ansatz for the time-evolving density matrix $\rho_A(t)$, which we numerically verified to reproduce the dynamics of entanglement with reasonable accuracy. 

Our results suggest some natural future directions. First, we have demonstrated that entanglement transitions can occur in generic boundary-driven Lindbladian dynamics. While transport properties in this setting have been extensively studied, entanglement dynamics remains largely unexplored. Second, the dynamics we studied can be implemented at scales intractable for classical simulations using platforms such as trapped ions or superconducting qubit arrays. In the latter case, boundary dissipation was recently realized using a single ancilla qubit~\cite{mi2024stable}. In these settings, the entanglement spectrum can be reconstructed using the ansatz in Eq.~\eqref{rhoansatz}, with free parameters estimated via the randomized measurement toolbox~\cite{elben2023randomized, joshi2023exploring, kokail2021entanglement, brydges2019probing, PhysRevLett.124.240505}. Third, while our numerical evidence explicitly demonstrates a transition in the min-entropy, our arguments suggest that other sufficiently large R\'enyi entropies should also exhibit a sharp transition, provided the Hamiltonian undergoes a finite-temperature phase transition. In this case, after the Page time, the R\'enyi entropy $S_\alpha$ is related to the equilibrium free energy at inverse temperature $\alpha$. When $\alpha$ is sufficiently large, the corresponding temperature lies within the symmetry-broken phase---which is macroscopically distinct from the state before the Page time. Studying these finite-$\alpha$ Page transitions requires sufficiently local dynamics (to ensure well-defined entanglement transitions) that also supports finite-temperature phase transitions. This can be achieved using two-dimensional geometries, such as in superconducting qubit arrays, or one-dimensional systems with long-range interactions falling off as $1/r^\beta$,  with $1 < \beta \leq 2$, as realized in trapped ion systems. Both approaches are experimentally feasible on present-day quantum hardware and can reach system sizes that are too large to simulate on classical computers.

One tantalizing perspective is to explore potential connections of such finite-$\alpha$ Page transitions to the island formula~\cite{Almheiri2019,Almheiri2020_1,Almheiri2020_2} in black hole physics, where the non-analytic emergence of a quantum extremal surface leads to non-analytic behavior of the von Neumann ($\alpha=1$) entanglement entropy. The island formula framework is currently regarded as the leading candidate for resolving the black hole information paradox.

\emph{Note added}.---Recently, a paper by Ganguly \textit{et al.}~\cite{ganguly2025quantum} appeared, addressing related but distinct questions. Our conclusions are consistent where they overlap. Additionally, a paper by Jha \textit{et al.}~\cite{RJha2025} studying a complementary model with charge conservation is scheduled to appear.

\subsection*{Acknowledgments}
We are grateful to David Huse and Achilleas Lazarides for suggesting the Lindbladian model. We are also grateful to Manoj Joshi and Peter Zoller for insightful discussions. L.H.L.~was supported by the National Science Foundation under Grant No.~DGE-2039656. S.K.~was supported by the Deutsche Forschungsgemeinschaft (DFG, German Research Foundation)-217133147/SFB 1073 (Project No.~B07). His work was performed in part at the Aspen Center for Physics, which is supported by National Science Foundation grant PHY-2210452, and at the Kavli Institute for Theoretical Physics (KITP), supported by grants NSF PHY-1748958 and PHY-2309135. S.G.~was supported through the Co-design Center for Quantum Advantage (C2QA) under contract number DE-SC0012704.
Any opinions, findings, and conclusions or recommendations expressed in this material are those of the authors and do not necessarily reflect the views of the National Science Foundation. \red{The data generated in this study are available from the corresponding author upon reasonable request.}

\balance
\bibliography{sources}

@article{kim2013ballistic,
  title={Ballistic spreading of entanglement in a diffusive nonintegrable system},
  author={Kim, Hyungwon and Huse, David A},
  journal={Physical review letters},
  volume={111},
  number={12},
  pages={127205},
  year={2013},
  publisher={APS},
  url={https://doi.org/10.1103/PhysRevLett.111.127205}
}

@incollection{polchinski2017black,
  title={The black hole information problem},
  author={Polchinski, Joseph},
  booktitle={New Frontiers in Fields and Strings: TASI 2015 Proceedings of the 2015 Theoretical Advanced Study Institute in Elementary Particle Physics},
  pages={353--397},
  year={2017},
  publisher={World Scientific},
  url={https://doi.org/10.1142/9789813149441_0006}
}

@article{almheiri2013black,
  title={Black holes: complementarity or firewalls?},
  author={Almheiri, Ahmed and Marolf, Donald and Polchinski, Joseph and Sully, James},
  journal={Journal of High Energy Physics},
  volume={2013},
  number={2},
  pages={1--20},
  year={2013},
  publisher={Springer},
  url={https://doi.org/10.1007/JHEP02(2013)062}
}

@article{akers2024black,
  title={The black hole interior from non-isometric codes and complexity},
  author={Akers, Chris and Engelhardt, Netta and Harlow, Daniel and Penington, Geoff and Vardhan, Shreya},
  journal={Journal of High Energy Physics},
  volume={2024},
  number={6},
  pages={1--119},
  year={2024},
  publisher={Springer},
  url={https://doi.org/10.1007/JHEP06(2024)155}
}

@article{kim2023complementarity,
  title={Complementarity and the unitarity of the black hole {S}-matrix},
  author={Kim, Isaac H and Preskill, John},
  journal={Journal of High Energy Physics},
  volume={2023},
  number={2},
  pages={1--46},
  year={2023},
  publisher={Springer},
  url={https://doi.org/10.1007/JHEP02(2023)233}
}

@article{dewolfe2023non,
  title={Non-isometric codes for the black hole interior from fundamental and effective dynamics},
  author={DeWolfe, Oliver and Higginbotham, Kenneth},
  journal={Journal of High Energy Physics},
  volume={2023},
  number={9},
  pages={1--22},
  year={2023},
  publisher={Springer},
  url={https://doi.org/10.1007/JHEP09(2023)068}
}

@article{PhysRevD.109.044005,
  title = {Information retrieval from {Hawking} radiation in the non-isometric model of black hole interior: Theory and quantum simulation},
  author = {Li, Ran and Wang, Xuanhua and Zhang, Kun and Wang, Jin},
  journal = {Phys. Rev. D},
  volume = {109},
  issue = {4},
  pages = {044005},
  numpages = {28},
  year = {2024},
  month = {Feb},
  publisher = {American Physical Society},
  doi = {10.1103/PhysRevD.109.044005},
  url={https://doi.org/10.1103/PhysRevD.109.044005}
}

@article{engelhardt2024cryptographic,
  title={Cryptographic censorship},
  author={Engelhardt, Netta and Folkestad, {\AA}smund and Levine, Adam and Verheijden, Evita and Yang, Lisa},
  journal={Journal of High Energy Physics},
  volume={2025},
  number={1},
  pages={1--58},
  year={2025},
  publisher={Springer},
  url={https://doi.org/10.1007/JHEP01(2025)122}
}

@article{harlow2013quantum,
  title={Quantum computation vs. firewalls},
  author={Harlow, Daniel and Hayden, Patrick},
  journal={Journal of High Energy Physics},
  volume={2013},
  number={6},
  pages={1--56},
  year={2013},
  publisher={Springer},
  url={http://doi.org/10.1007/JHEP06(2013)085}
}

@article{susskind2013black,
  title={Black hole complementarity and the {Harlow}-{Hayden} conjecture},
  author={Susskind, Leonard},
  journal={arXiv preprint arXiv:1301.4505},
  year={2013},
  url={
https://doi.org/10.48550/arXiv.1301.4505}
}

@article{page2013time,
  title={Time dependence of {Hawking} radiation entropy},
  author={Page, Don N},
  journal={Journal of Cosmology and Astroparticle Physics},
  volume={2013},
  number={09},
  pages={028},
  year={2013},
  publisher={IOP Publishing},
  url={https://doi.org/10.1088/1475-7516/2013/09/028}
}

@article{PhysRevD.100.105010,
  title = {Typical entanglement entropy in the presence of a center: {Page} curve and its variance},
  author = {Bianchi, Eugenio and Don\`a, Pietro},
  journal = {Phys. Rev. D},
  volume = {100},
  issue = {10},
  pages = {105010},
  numpages = {13},
  year = {2019},
  month = {Nov},
  publisher = {American Physical Society},
  url={https://doi.org/10.1103/PhysRevD.100.105010}
}

@article{blake2023page,
  title={The {Page} curve from the entanglement membrane},
  author={Blake, Mike and Thompson, Anthony P},
  journal={Journal of High Energy Physics},
  volume={2023},
  number={11},
  pages={1--28},
  year={2023},
  publisher={Springer},
  url={https://doi.org/10.1007/JHEP11(2023)016}
}

@article{PhysRevD.109.L081901,
  title = {Page-curve-like entanglement dynamics in open quantum systems},
  author = {Glatthard, Jonas},
  journal = {Phys. Rev. D},
  volume = {109},
  issue = {8},
  pages = {L081901},
  numpages = {7},
  year = {2024},
  month = {Apr},
  publisher = {American Physical Society},
  doi = {10.1103/PhysRevD.109.L081901},
  url = {https://doi.org/10.1103/PhysRevD.109.L081901}
}

@article{PhysRevD.102.086017,
  title = {Toy model for decoherence in the black hole information problem},
  author = {Agarwal, Kartiek and Bao, Ning},
  journal = {Phys. Rev. D},
  volume = {102},
  issue = {8},
  pages = {086017},
  numpages = {7},
  year = {2020},
  month = {Oct},
  publisher = {American Physical Society},
  doi = {10.1103/PhysRevD.102.086017},
  url = {https://doi.org/10.1103/PhysRevD.102.086017}
}

@article{piroli2020random,
  title={A random unitary circuit model for black hole evaporation},
  author={Piroli, Lorenzo and S{\"u}nderhauf, Christoph and Qi, Xiao-Liang},
  journal={Journal of High Energy Physics},
  volume={2020},
  number={4},
  pages={1--35},
  year={2020},
  publisher={Springer},
  url={https://doi.org/10.1007/JHEP04(2020)063}
}

@article{PhysRevB.109.224308,
  title = {Page curve entanglement dynamics in an analytically solvable model},
  author = {Kehrein, Stefan},
  journal = {Phys. Rev. B},
  volume = {109},
  issue = {22},
  pages = {224308},
  numpages = {11},
  year = {2024},
  month = {Jun},
  publisher = {American Physical Society},
  doi = {10.1103/PhysRevB.109.224308},
  url = { https://doi.org/10.1103/PhysRevB.109.224308}
}

@article{ganguly2025quantum,
  title={Quantum trajectories and {Page}-curve entanglement dynamics},
  author={Ganguly, Katha and Gopalakrishnan, Preethi and Naik, Atharva and Agarwalla, Bijay Kumar and Kulkarni, Manas},
  journal={arXiv preprint arXiv:2501.12110},
  year={2025},
  url={
https://doi.org/10.48550/arXiv.2501.12110}
}

@article{PhysRevLett.101.010504,
  title = {Entanglement Spectrum as a Generalization of Entanglement Entropy: Identification of Topological Order in Non-Abelian Fractional Quantum {Hall} Effect States},
  author = {Li, Hui and Haldane, F. D. M.},
  journal = {Phys. Rev. Lett.},
  volume = {101},
  issue = {1},
  pages = {010504},
  numpages = {4},
  year = {2008},
  month = {Jul},
  publisher = {American Physical Society},
  doi = {10.1103/PhysRevLett.101.010504},
  url = {https://doi.org/10.1103/PhysRevLett.101.010504}
}

@article{RevModPhys.83.863,
  title = {Colloquium: Nonequilibrium dynamics of closed interacting quantum systems},
  author = {Polkovnikov, Anatoli and Sengupta, Krishnendu and Silva, Alessandro and Vengalattore, Mukund},
  journal = {Rev. Mod. Phys.},
  volume = {83},
  issue = {3},
  pages = {863--883},
  numpages = {0},
  year = {2011},
  month = {Aug},
  publisher = {American Physical Society},
  doi = {10.1103/RevModPhys.83.863},
  url = {https://doi.org/10.1103/RevModPhys.83.863}
}

@article{PhysRevX.8.021026,
  title = {Does a Single Eigenstate Encode the Full {Hamiltonian?}},
  author = {Garrison, James R. and Grover, Tarun},
  journal = {Phys. Rev. X},
  volume = {8},
  issue = {2},
  pages = {021026},
  numpages = {24},
  year = {2018},
  month = {Apr},
  publisher = {American Physical Society},
  doi = {10.1103/PhysRevX.8.021026},
  url = {https://doi.org/10.1103/PhysRevX.8.021026}
}

@article{mi2024stable,
  title={Stable quantum-correlated many-body states through engineered dissipation},
  author={Mi, Xiao and Michailidis, AA and Shabani, Sara and Miao, KC and Klimov, PV and Lloyd, J and Rosenberg, E and Acharya, R and Aleiner, I and Andersen, TI and others},
  journal={Science},
  volume={383},
  number={6689},
  pages={1332--1337},
  year={2024},
  publisher={American Association for the Advancement of Science},
  url={https://doi.org/10.1126/science.adh9932}
}

@article{elben2023randomized,
  title={The randomized measurement toolbox},
  author={Elben, Andreas and Flammia, Steven T and Huang, Hsin-Yuan and Kueng, Richard and Preskill, John and Vermersch, Beno{\^\i}t and Zoller, Peter},
  journal={Nature Reviews Physics},
  volume={5},
  number={1},
  pages={9--24},
  year={2023},
  publisher={Nature Publishing Group UK London},
  url={https://doi.org/10.1038/s42254-022-00535-2}
}

@article{joshi2023exploring,
  title={Exploring large-scale entanglement in quantum simulation},
  author={Joshi, Manoj K and Kokail, Christian and van Bijnen, Rick and Kranzl, Florian and Zache, Torsten V and Blatt, Rainer and Roos, Christian F and Zoller, Peter},
  journal={Nature},
  volume={624},
  number={7992},
  pages={539--544},
  year={2023},
  publisher={Nature Publishing Group UK London},
  url={https://doi.org/10.1038/s41586-023-06768-0}
}

@article{brydges2019probing,
  title={Probing {R{\'e}nyi} entanglement entropy via randomized measurements},
  author={Brydges, Tiff and Elben, Andreas and Jurcevic, Petar and Vermersch, Beno{\^\i}t and Maier, Christine and Lanyon, Ben P and Zoller, Peter and Blatt, Rainer and Roos, Christian F},
  journal={Science},
  volume={364},
  number={6437},
  pages={260--263},
  year={2019},
  publisher={American Association for the Advancement of Science},
  url={https://doi.org/10.1126/science.aau4963}
}

@article{PhysRevLett.124.240505,
  title = {Quantum Information Scrambling in a Trapped-Ion Quantum Simulator with Tunable Range Interactions},
  author = {Joshi, Manoj K. and Elben, Andreas and Vermersch, Beno\^{\i}t and Brydges, Tiff and Maier, Christine and Zoller, Peter and Blatt, Rainer and Roos, Christian F.},
  journal = {Phys. Rev. Lett.},
  volume = {124},
  issue = {24},
  pages = {240505},
  numpages = {6},
  year = {2020},
  month = {Jun},
  publisher = {American Physical Society},
  doi = {10.1103/PhysRevLett.124.240505},
  url = {https://doi.org/10.1103/PhysRevLett.124.240505}
}

@article{kokail2021entanglement,
  title={Entanglement {Hamiltonian} tomography in quantum simulation},
  author={Kokail, Christian and van Bijnen, Rick and Elben, Andreas and Vermersch, Beno{\^\i}t and Zoller, Peter},
  journal={Nature Physics},
  volume={17},
  number={8},
  pages={936--942},
  year={2021},
  publisher={Nature Publishing Group UK London},
  url={https://doi.org/10.1038/s41567-021-01260-w}
}

@article{bulchandani2021superdiffusion,
  title={Superdiffusion in spin chains},
  author={Bulchandani, Vir B and Gopalakrishnan, Sarang and Ilievski, Enej},
  journal={Journal of Statistical Mechanics: Theory and Experiment},
  volume={2021},
  number={8},
  pages={084001},
  year={2021},
  publisher={IOP Publishing},
  url={https://doi.org/10.1088/1742-5468/ac12c7}
}

@book{fukunaga2013introduction,
  title={Introduction to statistical pattern recognition},
  author={Fukunaga, Keinosuke},
  year={2013},
  publisher={Elsevier}
}

@article{zhu2020entanglement,
  title={Entanglement {Hamiltonian} of many-body dynamics in strongly correlated systems},
  author={Zhu, W and Huang, Zhoushen and He, Yin-Chen and Wen, Xueda},
  journal={Physical review letters},
  volume={124},
  number={10},
  pages={100605},
  year={2020},
  publisher={APS},
  url={https://doi.org/10.1103/PhysRevLett.124.100605}
}

@article{RJha2025,
  title={Page Curve and Entanglement Dynamics in an Interacting Fermionic Chain},
  author={Jha, Rishabh and Manmana, Salvatore R and Kehrein, Stefan},
  journal={arXiv preprint arXiv:2502.03563},
  year={2025},
  url={
https://doi.org/10.48550/arXiv.2502.03563}
}

@article{Almheiri2019,
title={The entropy of bulk quantum fields and the entanglement wedge of an evaporating black hole},
author={Almheiri, A. and Engelhardt, N. and Marolf, D. and Maxfield, H.},
journal={J. High Energy Phys.},
year={2019},
volume={12}, 
pages={1},
url={https://doi.org/10.1007/JHEP12(2019)063}
}

@article{Almheiri2020_1,
title={Replica wormholes and the entropy of {Hawking} radiation},
author={Almheiri, A. and Hartman, T. and Maldacena, J. and Shaghoulian, E.
and Tajdini, A.}, 
journal={J. High Energy Phys.},
year={2020},
volume={5}, 
pages={1},
url={https://doi.org/10.1007/JHEP05(2020)013}
}

@article{Almheiri2020_2,
title={The {Page} curve of Hawking radiation from semiclassical geometry},
author={Almheiri, A. and Mahajan, R. and Maldacena, J. and Zhao, Y.},
journal={J. High Energy Phys.},
year={2020},
volume={3}, 
pages={1},
url={https://doi.org/10.1007/JHEP03(2020)149}
}

@article{saha2024generalized,
  title={Generalized Hydrodynamic Description of the {Page} Curve--like Dynamics of a Freely Expanding Fermionic Gas},
  author={Saha, Madhumita and Kulkarni, Manas and Dhar, Abhishek},
  journal={Physical Review Letters},
  volume={133},
  number={23},
  pages={230402},
  year={2024},
  publisher={APS},
  url={https://doi.org/10.1103/PhysRevLett.133.230402}
}

@article{glatthard2025thermodynamics,
  title={Thermodynamics of the {Page} curve in {Markovian} open quantum systems},
  author={Glatthard, Jonas},
  journal={arXiv preprint arXiv:2501.09082},
  year={2025},
  url={
https://doi.org/10.48550/arXiv.2501.09082}
}

@article{carlen2025stationary,
  title={Stationary states of boundary-driven quantum systems: {Some} exact results},
  author={Carlen, Eric A and Huse, David A and Lebowitz, Joel L},
  journal={Physical Review A},
  volume={111},
  number={1},
  pages={012210},
  year={2025},
  publisher={APS},
  url={https://doi.org/10.1103/PhysRevA.111.012210}
}

@misc{suppmat,
  title        = {See {Supplemental} {Materials}},
  note         = {Supporting numerical evidence of the Page curve and claims in the main text}
}


\clearpage

\setcounter{equation}{0}
\setcounter{figure}{0}
\setcounter{table}{0}
\setcounter{section}{0}
\setcounter{page}{1}
\makeatletter
\renewcommand{\theequation}{S\arabic{equation}}
\renewcommand{\thefigure}{S\arabic{figure}}
\renewcommand{\thepage}{S\arabic{page}} 
\renewcommand{\thesection}{S\arabic{section}} 
\renewcommand{\bibnumfmt}[1]{[S#1]}
\renewcommand{\citenumfont}[1]{S#1}

\title{Supplemental Materials: Sharp Page transitions in generic Hamiltonian dynamics}

\maketitle
\onecolumngrid
In this supplemental document we provide additional supporting numerical evidence for the claims in the main text: (\ref{sec:suppsec1}) we show how the Page curve depends on the size of the subsystem and its complement; (\ref{sec:flows}) we show how the energy profiles in the leading eigenvalues of the entanglement Hamiltonian evolve with time; (\ref{sec:suppsec3})~we show that the Page curve dynamics is qualitatively similar even if one does not initialize subsystem $A$ precisely in the ceiling state; (\ref{sec:suppsec4})~we present fits for a ``purely'' thermal ansatz for the reduced density matrix, not including the nonequilibrium corrections in the full ansatz in the main text; and (\ref{sec:recon})~we show that (at least for the full-system dynamics) the temperature-dependent thermal-state ansatz for the reduced density matrix accurately reproduces the entanglement entropies and the Page time.

\section{Effects of system size on the Page curve} \label{sec:suppsec1}

\begin{figure*}[h!]
    \centering
    \includegraphics{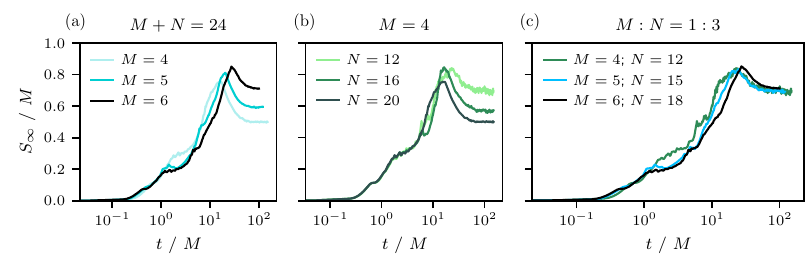}
    \caption{\red{Effects of system size}, keeping constant (a) the size of total system (subsystem plus the bath), (b) the size of the subsystem, and (c) the ratio between the size of the subsystem and the bath.}
    \label{fig:supp1}
\end{figure*}

Here, we show how the Page curve changes with the sizes $M$ and $N$ of the subsystem and its complement. First, we consider keeping $M + N$ constant (Fig.~\ref{fig:supp1}a). We see that the Page time increases with $M$, though we do not have enough dynamic range to tell how precisely it varies. \red{The min-entropy plateaus at a value that scales linearly with $M$; this reflects the amount of energy that was initially in the subsystem.} If instead the size of the subsystem is maintained at $M=4$ and the size of the bath is increased, the Page time remains approximately constant, but the final min-entropy value decreases with increasing bath size $N$ (Fig.~\ref{fig:supp1}b). This is expected since the entropy density of the final state decreases as $N$ is increased. Finally, when we keep the size of the subsystem and bath at a constant $1:3$ ratio and increase the total system size, we see that the Page curve stays roughly the same (Fig.~\ref{fig:supp1}c) with finite size effects being ironed out. In each of these cases, a clear Page curve is observed, with entanglement entropy increasing up to a maximum and subsequently decreasing. 

\begin{figure*}[h!]
    \centering
    \includegraphics{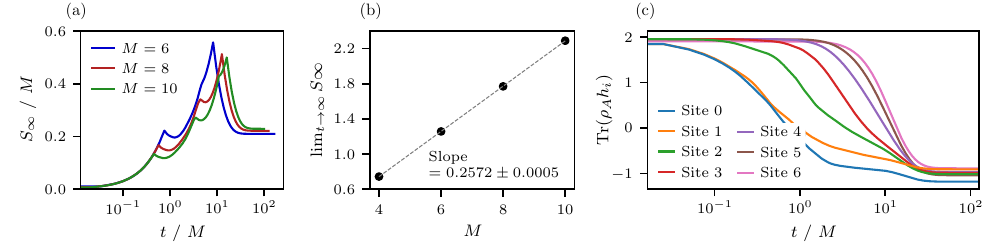}
    \caption{\red{Effects of system size on Lindbladian dynamics. We observe the (a) min-entropy $S_{\infty}$ for different system sizes; (b) volume law scaling of late-time $S_{\infty}$; and (c) local energy flows showing a nonequilibrium steady state for $M=8$.}}
    \label{fig:supp1-2}
\end{figure*}

\red{For Lindbladian dynamics, the short-time behavior remains consistent across different subsystem sizes (Fig.~\ref{fig:supp1-2}a), reflecting the fact that as the system cools, each larger subsystem must undergo the same cooling as its smaller counterparts. Although our jump operators always drive the edge spin to the ground state, the Hamiltonian is not frustration-free. Consequently, the late-time entanglement entropy scales linearly with the subsystem size $M$ (Fig.~\ref{fig:supp1-2}b). This is corroborated by the local energy distribution (Fig.~\ref{fig:supp1-2}c): energy remains present at all sites except site 0, consistent with the jump operators defined in terms of the local energy operator at sites 0 and 1.}

\section{Contributions to energy flow} \label{sec:flows}

\begin{figure*}[h!]
    \centering
    \includegraphics[width=475pt]{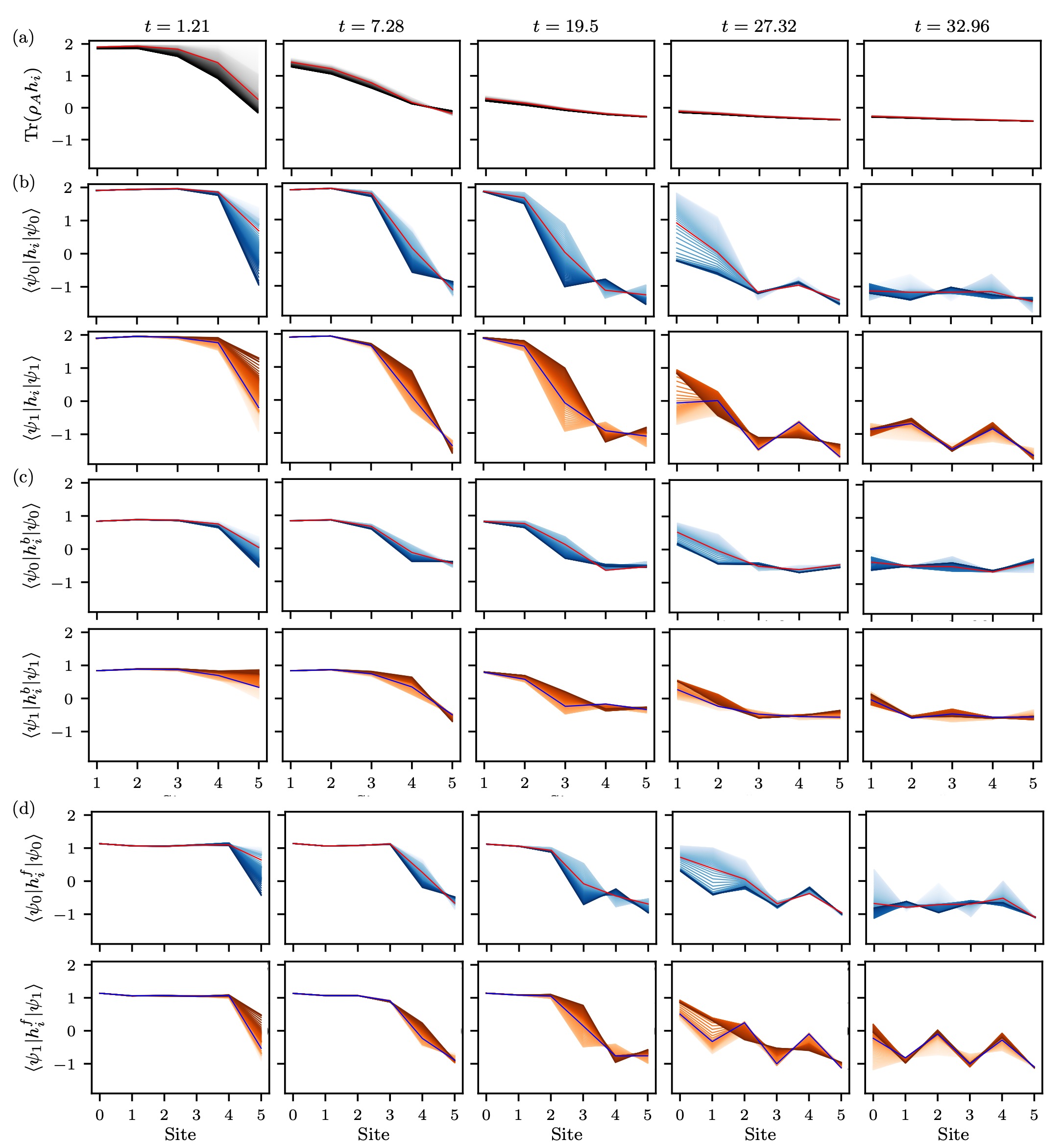}
    \caption{Contributions to energy flow out of the system. We show (a) the total energy profile $\Tr{\rho_A h_i}$; and the contributions from the top two eigenvectors $\ket{\psi_0}$ and $\ket{\psi_1}$ to the (b)  local energy $h_i$, (c) local bond energy $h_i^b$, and (d) local field energy $h_i^f$. The highlighted line indicates the energy profile at the given time, while light (dark) blue lines indicate a succession of previous (future) times.}
    \label{fig:supp3}
\end{figure*}

In Fig.~\ref{fig:supp3}, we present the time-evolving local energy profile of the density matrix $\rho_A(t)$ and its leading few eigenvectors at times before (and approaching) the Page time. The local energy of the full density matrix is given by $\Tr{\rho_A h_i}$ (Fig.~\ref{fig:supp3}a), where $\rho_A$ is the reduced density matrix of the system and $h_i$ is the local energy operator. We may calculate the contribution of the top two eigenvectors of $\rho_A$, $\ket{\psi_0}$ and $\ket{\psi_1}$ to the local energy $h_i$ (Fig.~\ref{fig:supp3}b), local bond energy $h_i^b = J\sigma_i^z\sigma_{i+1}^z$ (Fig.~\ref{fig:supp3}c), and local field energy $h_i^f = g\sigma_i^x + h\sigma_i^z$ (Fig.~\ref{fig:supp3}d). Each of the shown times $t$ mark avoided crossings between eigenvalues, and we see Landau-Zener-like swapping of the eigenvectors. We observe that the total energy dips below the zero at the Page-time ($t=27.32$).

\section{Initial states away from the top of the spectrum} \label{sec:suppsec3}

\begin{figure*}[h!]
    \centering
    \includegraphics{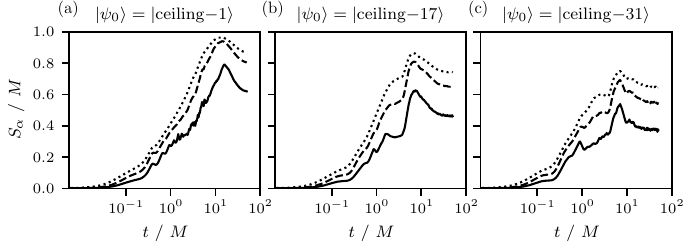}
    \vspace{-8pt}
    \caption{Renyi entropies where the subsystem is not initialized in the ceiling state, but rather in the (a) 2nd, (b) 18th, and (c) 32nd most excited state. In all cases, evidence of a Page curve is still present.}
    \label{fig:supp4}
\end{figure*}

In the main text, we discussed dynamics starting from a state in which subsystem $A$ is initialized in the top eigenstate of $H$. A natural question is whether this physics depends on starting precisely in this state. As documented in this section, we still observe a Page-curve, namely the downbending of entanglement entropy, when we initialize the system in a lower-energy excited state. As we initialize the state in the 2nd (Fig.~\ref{fig:supp4}a), 18th (Fig.~\ref{fig:supp4}b), and 32nd (Fig.~\ref{fig:supp4}c) state, counting from the top of the spectrum, we see that the signal for the Page curve gets weaker, but persists. In general, we also find that the Page time is shorter when the system is initialized in a less excited eigenstate.

\section{$\beta$ fits with and without current terms}
\label{sec:suppsec4}

\begin{figure*}[h!]
    \centering
    \includegraphics[width=500pt]{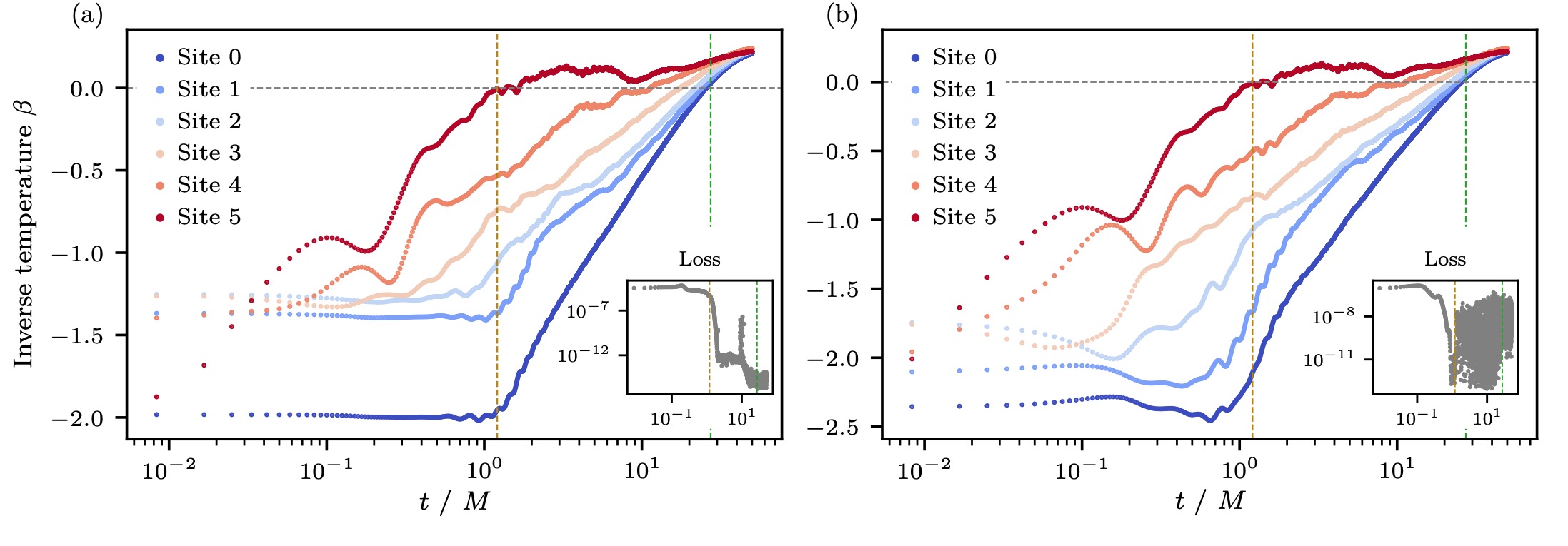}
    \caption{Fits to a spatially-dependent temperature (a) with and (b) without  extra momentum terms. These are full-dynamics simulations.}
    \label{fig:supp7}
\end{figure*}

In the main text, we add terms to our ansatz representing the current flowing out of the subsystem $A$, to allow for more fitting parameters (Fig.~\ref{fig:supp7}a). In the case that we do not include these extra terms, the error in our fits increases (Fig.~\ref{fig:supp7}b). Nevertheless, the fitted values for $\beta$ still show the qualitative features we expect.

\section{Reconstruction of Renyi entropies}
\label{sec:recon}
\begin{figure*}[h!]
    \centering
    \includegraphics{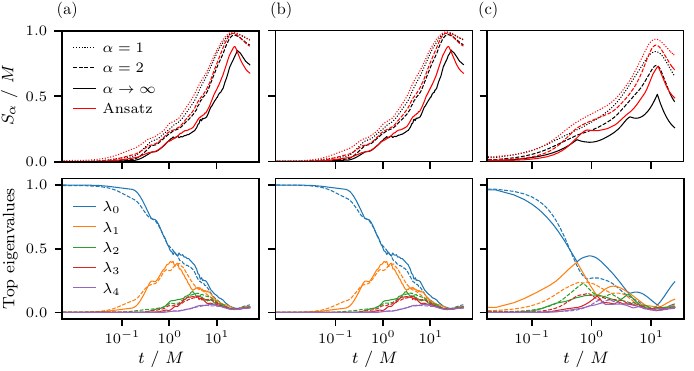}
    \caption{Reconstructed Renyi entropies and eigenvalue spectrum from entanglement Hamiltonian ansatz for full-system dynamics (a) with and (b) without momentum terms, and (c) Lindblad dynamics. Dashed lines indicate the reconstructed eigenvalue spectrum.}
    \label{fig:supp5}
\end{figure*}

We demonstrate that our ansatz is indeeed a good approximation for the reduced density matrix of the system in both full-system dynamics with and without momentum terms (Figs.~\ref{fig:supp5}a,b) and Lindbladian dynamics (Fig.~\ref{fig:supp5}c). That is, we can use the entanglement Hamiltonian ansatz with spatially-dependent temperature to reconstruct the Renyi entropies and spectrum of our system. In both cases the Ans\"atze correctly capture the Page time. In the case of Lindbladian evolution, the match between the Ansatz and the true dynamics is systematically worse. We attribute this to the fact that the energy current out of the system remains large across the Page time, making a hydrodynamic ansatz less successful. 

\end{document}